# Digesting Network Traffic for Forensic Investigation Using Digital Signal Processing Techniques

S. Mohammad Hosseini, Amir Hossein Jahangir, Mehdi Kazemi

*Abstract*— One of the most important practices of cybercrime investigations is to search a network traffic history for an excerpt of traffic, such as the disclosed information of an organization or a worm's signature. In post-mortem investigations, criminals and targets are detected by attributing the excerpt to payloads of traffic flows. Since it is impossible to store the high volume of data related to long-term traffic history, payload attribution systems (PAS) based on storing a compact digest of traffic using Bloom filters have been presented in the literature. In these systems, querying the stored digest for an excerpt results in a low number of suspects instead of certain criminals. In this paper, we present a different PAS which is based on simple and widespread digital signal processing techniques. Our traffic digesting scheme has been inspired by DSP-based compression algorithms. The proposed payload attribution system, named DSPAS, not only results in a low false positive rate but also outperforms previous schemes in response to wildcard queries which are essentially applicable for complex excerpts such as the signature of polymorphic worms. Our theoretical analysis and practical evaluations show that DSPAS results in a significantly lower false positive rate and also a lower processing time for wildcard queries in comparison to previous works. Moreover, our PAS can prevent a malicious insider from evading the PAS by its ability to find strings similar to a queried excerpt.

*Index Terms*— Payload attribution system, Cybercrime investigation, Transform coding, Traffic digesting.

## I. Introduction

Computer networks are always under the threat of cybercrimes. No security system can provide absolute security. Even best network security systems cannot detect and prevent all attacks, especially new and unknown ones. In some cases, it is impossible to prevent cybercrimes. For example, if confidential information of an organization leaks via its network, how can security experts find the cybercriminal? To give another illustration, assume a worm has spread in the internal network of an organization, and the Intrusion Detection and Prevention System (IDS/IPS) of the organization could not detect and prevent the worm. How is it possible to find the individual who spread it or to find the infected systems? Hence, it is necessary to have tools and techniques besides preventive security systems to investigate cybercrimes after their occurrence. This is the role of network forensics and its tools.

The simplest approach to network investigation is to record and store raw network traffic [1], [2]. It is possible to investigate any network incident by traffic recording. An investigator can search the recorded traffic for the disclosed information or the worm's signature to find its source and destination. This procedure is called "attribution." The most challenging problem of this scheme is the storage of the high volume of data which is extremely expensive [3]. The other problem with traffic recording is the violation of privacy [4]. It is possible to access personal information of users by recording network traffic. Providing both privacy and network forensics has become so challenging that new architectures and protocols have been proposed for the Internet [5], [6], however, applying such changes is too expensive to make them practical.

To resolve these problems, Kulesh et al. [7], [8] have proposed the first payload attribution system (PAS) based on storing a compact digest of network traffic. The system, named Hierarchical Bloom Filter (HBF), digests traffic using hash functions and Bloom filters [9]. The digest, which is considerably smaller than the original traffic, cannot be reverted to the original data and therefore, privacy is preserved. However, HBF can respond to a query for an excerpt of traffic, i.e. it can detect the flows (source and destination pairs) that carried the excerpt. There is, however, a drawback with HBF which is its non-zero false positive rate. It means the results obtained by HBF are only suspects, not certain criminals. Network incident analysts use this investigative tool and the provided aids to find the criminals who committed a cybercrime and also to identify other victims of an attack.

Next studies have presented improved payload attribution systems in terms of false positive rate and data reduction ratio [10]–[13]. Nevertheless, an important problem that has not been adequately addressed by the previous works is wildcard queries. A wildcard query is a query for an excerpt in which some bytes are unknown, such as the signature of

The authors are with the Department of Computer Engineering, Sharif University of Technology, Tehran 145888-9694, Iran (e-mail: smhoseini@ce.sharif.ir; Corresponding author: jahangir@sharif.ir; mehdi.kazemi@alum.sharif.ir).

polymorphic worms. A polymorphic worm changes its appearance with each instance. However, previous studies on these worms indicate that some invariant parts can be extracted from the byte code of a polymorphic worm, and the parts are then separated by random bytes [14]. Hence the signature of interest for such a worm would be in the form of "A*B", where "A" and "B" are two invariant strings separated by a set of unknown random bytes. The previous payload attribution methods need to query for all possible values of the unknown bytes which not only takes an excessively long time but also imposes a high false positive rate because each simple query has a possibility of false positive response. It is noteworthy that even the intrusion detection systems that are specially designed for polymorphic worms could not achieve a false negative rate of zero [15]. Hence, it is essential to have a payload attribution system capable of processing wildcard queries for post-mortem investigations of such incidents.

Another important drawback of the previous works is that a malicious insider can easily evade a PAS by adding small changes in the transferred information. Suppose an insider wants to send out confidential information of an organization. The insider is aware of the deployed PAS based on digesting traffic using Bloom filters, i.e. as the methods used in the previous papers. They have a very high false positive rate for small excerpts (smaller than 200 bytes). Hence, the insider simply changes one byte of the information (or adds one byte) after every 200 bytes. As a result, querying for excerpts smaller than 200 bytes (if they do not contain the changed byte) results in a very high false positive rate. On the other hand, querying for excerpts of the information that are larger than 200 bytes results in a false negative response. Hence, a solution is needed that can query the digest for strings similar to the queried excerpt. As a result, an investigator can query for a longer excerpt (larger than 200 bytes) and find it in the digest even with some different bytes.

To resolve the problems of wildcard queries and querying for similar strings, we tried to seek other approaches instead of Bloom filter and hash functions, because the smallest change in the input of a hash function results in a completely different digest. We needed compression and data reduction techniques which are not very sensitive to small changes. This requirement conducted us to digital signal processing methods, especially transform coding. In this paper, we present a DSP-based Payload Attribution System, named DSPAS. Our approach, which uses simple and widespread DSP techniques, results in a significantly lower false positive rate for wildcard queries, while also reducing the response time. The approach can easily find strings in a digest that are similar to a queried excerpt. To the best of our knowledge, there is no similar study on network investigative methods and traffic archiving that uses digital signal processing techniques.

The rest of this paper is organized as follows: Section II briefly discusses related works. The proposed method is presented in Section III. In Section IV, the behavior of the proposed method is theoretically analyzed, and Section V evaluates the method. We conclude the paper in Section VI.

## II. RELATED WORKS

As mentioned before, all the previous methods of network traffic digesting are based on the Bloom filter. A Bloom filter [9] is a space-efficient probabilistic data structure that uses hash functions for storing the members of a set. A Bloom filter cannot present the members that it has stored, but it can answer membership queries with a deterministic false positive rate. Bloom filters have been widely used in many network applications [16]. However, since our approach to the traffic digesting problem is entirely different, we briefly review the previous works, and we do not go into details of them. For a comprehensive survey of the previous works, refer to [13].

HBF [7], which is the first traffic digesting scheme, partitions each packet of network traffic into equal-sized blocks and inserts the blocks into a Bloom filter after hashing. Moreover, a combination of each packet piece and the flow identifier of the packet is inserted into the Bloom filter. HBF stores a Bloom filter as the traffic digest of each period in a permanent storage system. For example, it may be a Bloom filter per hour. Furthermore, a list of all seen flows during each period is saved.

The investigation for an excerpt of traffic is done in two steps: In the first step, called "appearance check," the excerpt is partitioned into blocks, and the stored digest is searched for the blocks to determine the Bloom filter (time period) containing the excerpt. In the second step, called "flow determination," the excerpt blocks are combined with the flows of the determined Bloom filter and finally, searching for the combined blocks results in the flows that carried the excerpt. Checking the blocks alignment and the consecutiveness of blocks were important problems addressed by HBF.

Ponec et al. [10], [11] proposed the improved methods WBS and WMH which achieved significantly lower false positive rates at a data reduction ratio as high as 100:1. They used Winnowing fingerprinting technique [17] for partitioning a payload into blocks and inserting them into a Bloom filter. As a result, the blocks alignment problem was resolved more efficiently.

Haghighat et al. [12] proposed the CMBF method, which supports wildcard queries. As stated before, a wildcard query is a query for an excerpt in which some bytes are unknown. An example of a wildcard query can be an investigation for the signature "A????????B" which is composed of two invariant strings "A" and "B" with 8 unknown random bytes between them. The methods prior to CMBF require an exhaustive search, i.e. a simple query for each possible value of a wildcard excerpt, which is practically infeasible due to a large number of simple queries ($2^{64}$ simple queries for the above example). CMBF ameliorates this problem by classifying byte values and decreasing the number of required simple queries.

CBID is the most recent proposed PAS which is based on a combination of Bloom filter, Bitmap index table and a new traffic downsampling technique [13]. It outperforms the previous methods in terms of false positive rate. Nonetheless, CBID does not support wildcard queries, and therefore, it needs an exhaustive search for a wildcard query just like the methods prior to CMBF. However, even CMBF has important drawbacks regarding wildcard queries.

Although CMBF has significantly decreased the simple queries needed for a wildcard query, the achieved reduction in simple queries is not enough to keep the false positive rate low. Since each simple query has a possibility of a false positive response, the large number of them imposes a significantly large false positive rate for wildcard queries as shown in the evaluation section of this paper. It should be noted that paper [12] has not reported the false positive rate of wildcard queries. Moreover, the response time of CMBF increases exponentially with the number of wildcard bytes of an excerpt so that querying for an excerpt composed of more than eight wildcards would be infeasible.

All the problems associated with the performance of wildcard queries are due to the sensitivity of hash functions to their input. Using hash functions means that we do not tolerate even small differences and it makes us perform many queries for an excerpt composed of wildcard bytes. Hence, we sought a different approach for traffic digesting which can process wildcard queries just like the simple queries. It conducted us to an approach which can query a digest for strings similar to a queried excerpt. In the next section, we present our DSP-based payload attribution system which is a new scheme in the literature.

## III. DSPAS

The first requirement that must be addressed by a payload attribution system is data reduction. From this point of view, a procedure similar to the lossy compression algorithm JPEG, which is based on the transform coding technique, may be useful. In transform coding, a signal is transformed to, and expressed in, the frequency domain. Then, the frequency domain coefficients are quantized. Based on the granularity of the quantization which is coarser for less important coefficients, a portion of the coefficients is represented by fewer bits. As a result, a considerable number of the coefficients become zero, and therefore they will have low entropy. In the last step, the coefficients are compressed by entropy coding algorithms such as run-length encoding and LZMA [18]. The inverse of this procedure yields a signal similar to the original signal. The level of similarity between the original signal and the retrieved signal is dependent on the quantization granularity.

The Discrete Cosine Transform (DCT) [19] is a commonly used transform for this lossy compression technique, especially for images [20], which results in a good representation of the original signal with a low number of frequency domain coefficients. Likewise, the DCT transform, which is computed for the signal $x_n$ with a length of $L$ using relation (1), can be a logical choice for our application.

$$X_k = \sum_{n=0}^{L-1} x_n \cos\left[\frac{\pi}{L}\left(n+\frac{1}{2}\right)k\right], \quad k = 0,1,\ldots,L-1 \quad (1)$$

We used this scheme to store an approximation of traffic payloads. One difference between the requirement of a PAS and JPEG is that a PAS does not need to store a signal as much like as the original signal. An approximation of the original signal with the ability to respond to a query for checking the presence of an excerpt is satisfactory for a PAS. Therefore, we can use a very coarse quantization and achieve a high data reduction ratio.

Fig. 1 shows the distribution of Discrete Cosine Transform (DCT) coefficients of the payload of a traffic flow. As can be seen, the coefficients have a Gaussian distribution with a mean of zero. Network traffic typically has high entropy, and its data bytes are, to a large extent, independent and uniform [21]. Therefore, the Gaussian distribution appears due to the Central Limit Theorem (CLT) [22]. The Gaussian distribution means that a small part of coefficients comprises a considerable part of the signal's energy, and consequently, we can use the high energy coefficients as the digest of each payload. Hence, our solution for digesting the payload of a traffic flow is to remove the low energy coefficients and store an approximation of a payload by saving only the high energy coefficients.

Fig. 2 represents our approach to traffic digesting. Unlike the previous methods, DSPAS does not digest individual packet payloads, because the payload of a packet may be smaller than what we need for the DCT transformation. DSPAS digests the packet payloads of each traffic flow as a whole. In other words, we digest flow payloads instead of packet payloads. Therefore, in the first step, DSPAS classifies the input traffic and separates its flows. The traffic classification can be offloaded to an intrusion detection system (IDS) since an IDS typically does this task.

Then, DSPAS performs preprocessing operations on the payload of each flow. In this step, a $W$-byte window is slid through the payload and for each position of the window the hash of payload bytes inside the window is computed. The output of the hash function is a $W$-byte word, and the hash window is moved $W$ bytes in each step. The operations of the next steps are carried out on the hashed values. The purpose of hashing is to make values of payload bytes more independent and uniform. Hence, the DCT coefficients effectively tend to a Gaussian distribution. Moreover, hashing the payloads guarantees privacy protection. In addition to hashing, long strings of repetitive bytes are removed in the preprocessing step. These low entropy strings not only are useless for forensic purposes, but also prevent the DCT coefficients from becoming Gaussian. In the next step, the $W$-byte words are fed into the DCT unit. In the DCT transform step, the input signal is divided into chunks of size $L$ words and then transformed into the frequency domain using the DCT transform. Flows smaller than the transform size, $L$, (and also the last chunk of each flow) are padded with uniform random bytes before being transformed.

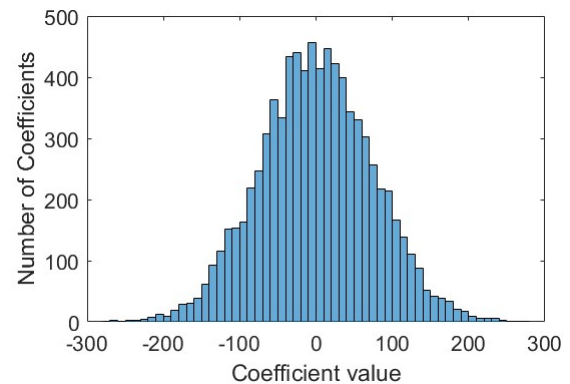

Fig. 1. Distribution of DCT coefficients of a payload

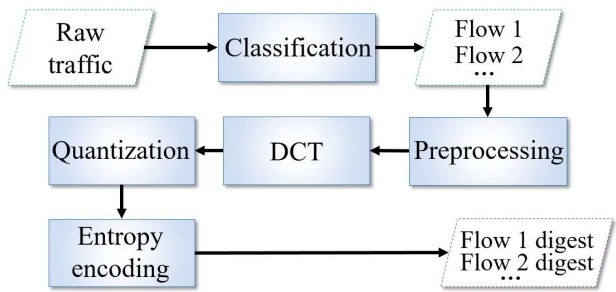

Fig. 2. Traffic digesting procedure of DSPAS

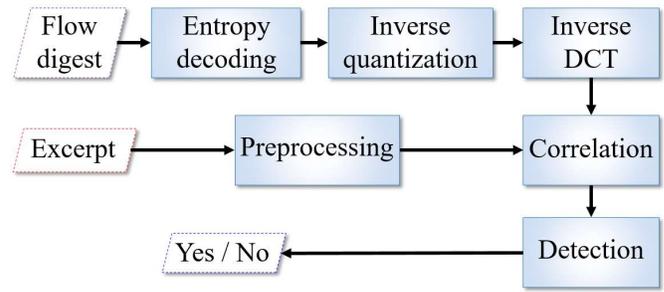

Fig. 3. Investigation procedure of DSPAS

In the next step, DSPAS quantizes the frequency domain coefficients by rounding them to the nearest integer value. Then, a certain number of least significant bits of the coefficients are removed, and only $q$ most significant bits of each coefficient are stored. The data reduction ratio after this step, $DR_q$, is equal to:

$$DR_q = W \times 8 / q \quad (2)$$

Since the coefficients have a Gaussian distribution with a mean of zero, many coefficients become zero after the quantization step. Hence, the output of this step is expected to have long strings of zeros and consequently, low entropy. In the last step, DSPAS compresses the data using a lossless entropy coding algorithm. The number of quantization bits ($q$) has a significant impact on the compression ratio of the entropy coding step. Setting a low value for $q$ can effectively decrease the coefficients entropy and consequently increases the compression ratio. However, a very low value of $q/W$ can drastically distort the signal.

The investigation procedure for an excerpt is shown in Fig. 3. First, DSPAS carries out the preprocessing operations on the excerpt. The entropy coding, the quantization, and the transformation are also inversely performed on the digest of each traffic flow. In the next step, DSPAS slides the excerpt through the retrieved payload and computes the cross-correlation between them. It is expected that a clear peak appears in the correlation signal if the excerpt exists in the payload. The flow of each payload that contains such a peak is determined as a carrier of the excerpt. The peak is detected by setting a threshold. As the excerpt becomes longer, the ratio of the peak value to the standard deviation of the correlation signal increases. Therefore, the threshold can be adaptively set according to the excerpt length in order to have a less false positive rate for larger excerpts. Hence, DSPAS uses this equation to set the threshold $T_l$ for excerpts of size $l$:

$$T_l = K_l\, \sigma_c \quad (3)$$

where $\sigma_c$ is the standard deviation of the correlation signal and $K_l$ is a predefined constant value for excerpts of size $l$. It should be noted that setting a proper threshold is a very important issue in this approach. A low value for the threshold may result in a high false positive rate. On the other hand, a high threshold value can result in false negatives. However, detecting the peak using a simple threshold based on the standard deviation is the most straightforward solution. One can use more sophisticated CFAR (Constant False Alarm Rate) techniques [23]. It should also be noted that DSPAS suffers from the alignment problem just like CMBF and due to this problem, the correlation must be computed using $W$ different alignments of the excerpt.

Regarding wildcard queries, DSPAS replaces each wildcard byte of the excerpt with the mean of all possible values, i.e. byte zero, and then, treats the modified excerpt just as simple (non-wildcard) queries. It is expected that the correlation signal is not significantly affected if the number of wildcard bytes is considerably lower than the excerpt length. Hence, the wildcard queries can also be simply detected in the same way as simple queries. This achievement is a major advantage of DSPAS. While CMBF needs to query for all possible values of the wildcard bytes, DSPAS uses only a single query for an excerpt composed of unknown bytes. Hence, DSPAS does not suffer from a high false positive rate for wildcard queries. Moreover, the correlation-based detection method gives the PAS the ability to find strings similar to a queried excerpt. Therefore, a malicious insider cannot easily evade the PAS as discussed in the introduction section.

In the next section, we model the performance of the proposed method by a theoretical analysis to clarify the impact of the system parameters.

## IV. THEORETICAL ANALYSIS

In this section, we theoretically analyze the proposed method and provide formulas for its false positive and false negative rates.

An additive noise on a payload can model the effect of digesting traffic using the proposed method. Therefore, we have:

$$P = I + N \quad (4)$$

where $I$ is the intact payload which is the input of the digesting system, $P$ is the recovered payload from its digest, and $N$ is the noise imposed by the digesting method. We experimentally studied recovered payloads from digested traffic and observed that their difference with original ones is always a white Gaussian noise, as it was naturally expected. For any fixed set of the PAS parameters, the system results in an additive white Gaussian noise (AWGN) with a mean of zero and a variance of $\sigma_N^2$. Moreover, the correlation signal, which is the summation of independent and identically distributed variables, tends to a Gaussian distribution according to Central Limit Theorem. It should also be noted that as discussed in the previous section, $P$ and $I$ follow the uniform distribution $U(-A, A)$[1] where the interval size depends on $W$. For example, $W = 1$ results in $A = 127$. Given these preliminaries, we analyze the false positive and false negative rates.

Assume $S$ be an excerpt of size $l$ that we are going to detect in the noisy payload $P$. The correlation signal, $C$, is calculated using the following relation:

$$C_n = \sum_{i=1}^{l} P_{n+i-1} S_i, \quad 1 \leq n \leq L - l + 1 \quad (5)$$

After determining the correlation signal, we set the threshold stated in relation (3) to detect the excerpt, i.e. the standard deviation of the correlation signal multiplied by a predefined coefficient ($K_l$). A false positive occurs when the correlation signal goes higher than the threshold at a point while the excerpt is not present at that point. Since the correlation signal has a Gaussian distribution, each point of $C_n$ crosses the threshold with a probability of $Q(K_l)$ according to Q-function analysis. $C_n$ is $L - l + 1$ points long and consequently, the number of false positives follows the binomial distribution $X \sim B(L - l + 1, Q(K_l))$. Therefore, the probability of false positive is:

$$\Pr(X > 0) = 1 - \Pr(X = 0)$$
$$= 1 - \binom{L - l + 1}{0} Q(K_l)^0 (1 - Q(K_l))^{L-l+1}$$
$$= 1 - (1 - Q(K_l))^{L-l+1} \quad (6)$$

Increasing the threshold ($K_l$) results in a lower false positive probability, which is desirable. However, when numerical techniques such as the correlation are used, we cannot discuss only the false positive probability without taking the false negative probability into account. The possibility of a false negative result is a drawback of our new approach which must be controlled and set as nearest as possible to zero. A false negative occurs if the correlation value of the point $n = Z$ at which the excerpt exists does not reach the threshold. The correlation value at the point is equal to:

$$C_{n=Z} = C_Z = \sum_i S_i P_{Z+i-1} = \sum_i S_i (S_i + N_i) \quad (7)$$

where $N$ is an additive white Gaussian noise (AWGN) with a variance of $\sigma_N^2$. Therefore, the false negative probability is determined by:

$$\Pr(C_Z < T) = \Pr(C_Z < K_l \sigma_{C_n}) = 1 - \Pr(C_Z > K_l \sigma_{C_n})$$
$$= 1 - \Pr\left(\frac{C_Z - \mu_{C_Z}}{\sigma_{C_Z}} > \frac{K_l \sigma_{C_n} - \mu_{C_Z}}{\sigma_{C_Z}}\right)$$
$$= 1 - Q\left(\frac{K_l \sigma_{C_n} - \mu_{C_Z}}{\sigma_{C_Z}}\right) \quad (8)$$

To achieve the false negative probability formula, we need to determine $\sigma_{C_n}$, $\mu_{C_Z}$ and $\sigma_{C_Z}$. For the standard deviations, first, we determine the corresponding variances:

$$\sigma_{C_n}^2 = Var\left(\sum_i P_i S_i\right) = Var\left(\sum_i (I_i + N_i) S_i\right)$$
$$= \sum_i Var(I_i S_i) + \sum_i Var(N_i S_i)$$
$$= \sum_i (\sigma_I^2 \sigma_S^2 + \sigma_I^2 \mu_S + \sigma_S^2 \mu_I)$$
$$+ \sum_i (\sigma_N^2 \sigma_S^2 + \sigma_N^2 \mu_S + \sigma_S^2 \mu_N) \quad (9)$$

$S$ and $I$ are uniformly distributed in the interval $[-A, A]$. Therefore:

$$\mu_I = \mu_S = 0 \quad (10)$$
$$\sigma_I^2 = \sigma_S^2 = \frac{(-A - A)^2}{12} = \frac{A^2}{3} \quad (11)$$
$$\sigma_{C_n}^2 = \sum_i \frac{A^4}{9} + \sum_i \frac{\sigma_N^2 A^2}{3}$$
$$= \frac{lA^4}{9} + \frac{l\sigma_N^2 A^2}{3}$$
$$\sigma_{C_n} = A\sqrt{\frac{lA^2}{9} + \frac{l\sigma_N^2}{3}} \quad (12)$$

Similarly, we determine $\sigma_{C_Z}$:

$$\sigma_{C_Z}^2 = Var\left(\sum_i S_i (S_i + N_i)\right)$$
$$= Var\left(\sum_i S_i N_i\right) = \sum_i Var(S_i N_i)$$

---

[1] The exact distribution is $U(-A-1, A)$. However, to keep it simple and straightforward, we used the notation $U(-A, A)$. The two notations are approximately equal for a large value of $W$.

$$= \sum_i \sigma_S^2 \sigma_N^2 = l\sigma_S^2 \sigma_N^2 = l\sigma_N^2(E(S^2) - E^2(S))$$

$$= l\sigma_N^2 \frac{\sum_i S_i^2}{l} = \sigma_N^2 \sum_i S_i^2 \quad (13)$$

$\sum_i S_i^2$ can be easily calculated using the variance of $S_i$ (Relation 11):

$$\sigma_S^2 = \frac{A^2}{3} = E(S^2) - E^2(S) = E(S^2)$$

$$= \frac{\sum_i S_i^2}{l}$$

$$\sum_i S_i^2 = \frac{lA^2}{3} \quad (14)$$

Therefore:

$$\sigma_{C_Z}^2 = \frac{\sigma_N^2 lA^2}{3}$$

$$\sigma_{C_Z} = A\sqrt{\frac{\sigma_N^2 l}{3}} \quad (15)$$

And the final term, $\mu_{C_Z}$:

$$\mu_{C_Z} = E\left(\sum_i S_i^2 - \sum_i S_i N_i\right) = \sum_i S_i^2 = \frac{lA^2}{3} \quad (16)$$

Based on relations 8, 12, 15 and 16, the false negative probability equals:

$$\Pr(C_Z < T) = 1 - Q\left(\frac{K_l A\sqrt{\frac{lA^2}{9} + \frac{l\sigma_N^2}{3}} - \frac{lA^2}{3}}{A\sqrt{\frac{\sigma_N^2 l}{3}}}\right) \quad (17)$$

Relation (17) represents the false negative probability as a function of the threshold coefficient ($K_l$), the excerpt size ($l$), the interval of the uniform distribution of input data ($A$) and the variance of the noise imposed by digesting ($\sigma_N^2$). Given the above analysis, we can discuss the system parameters' effects:

- Transform size ($L$): increasing transform size results in a higher false positive probability according to relation (6). On the other hand, a small transform size can increase the probability that a transform block does not entirely include a queried excerpt and consequently, the effective excerpt length ($l$) may be smaller than the excerpt's actual length. Hence, a very small transform size can also increase the false positive probability according to the same relation. Obviously, the optimum transform size is dependent on the excerpt size. In the next section, we practically evaluate this parameter.
- Correlation threshold coefficient ($K_l$): according to relation (6), we can achieve a lower false positive probability by increasing the threshold. On the other hand, relation (17) shows that increasing the threshold coefficient results in a higher false negative probability, which is undesirable. As relation (17) indicates, a larger excerpt size ($l$) can decrease the input value of the Q-function and compensate for the negative effect of increasing the threshold coefficient. It means that for a fixed threshold coefficient, the false negative probability of larger excerpts is lower. Consequently, it is possible to increase the threshold for larger excerpts and achieve lower false positive and false negative probabilities.
- Word size ($W$) and quantization bits ($q$): the combination of these two parameters directly affects the data reduction ratio, the imposed noise and the interval of the uniform distribution of the input data ($A$). To achieve the best processing speed, we use the larger word size that our CPU supports, i.e. $W = 8$ bytes (for the usual 64-bit CPUs). It results in $A \approx 9.2e18$. Given the 8-byte word size, different values of $q$ must be evaluated to obtain an acceptable result.

In the next section, we present our practical results and compare them with the results expected from the theoretical analysis. We also evaluate different values for the system parameters. It should be noted that similarly to the previous papers, we use the term "rate" instead of empirical probabilities determined by experiments.

## V. EVALUATION

In this section, first, we evaluate DSPAS parameters to determine proper values for them. Then, we compare DSPAS results with the previous works in response to both simple and wildcard queries. In addition to DSPAS, we implemented the methods WMH, CMBF, and CBID. We collected 4 GB of TCP and UDP traffic of our CE Department's core switch. The traffic, which was uniform at the recording time, comprises 76% TCP and 24% UDP flows. In the following, we evaluate $q$, $K_l$ and $L$, respectively. Then, we compare the false positive rate, false negative rate, and processing time of our method with the previous works.

In the first step, we need to determine the proper value for quantization bits ($q$). We digested the traffic trace by DSPAS using $W = 8$, $L = 1024$ and different values of $q$ (in the following we evaluate and discuss $L$). Table I shows the resulting data reduction ratio, the variance of the imposed noise, and the signal-to-noise ratio. Moreover, Fig. 4 represents examples of the correlation signal for each experiment. Although $q = 3$ results in a good data reduction ratio, its SNR is approximately equal to 1 which means the power of the imposed noise is equal to the signal's power. Hence, the recovered payload would be completely distorted. It can be clearly seen in Fig. 4 which shows an ineffective correlation signal for $q = 3$. On the other side, $q = 5$ achieves a high SNR. It results in a signal power approximately twice the noise power. As we show in the following, the SNR of $q = 4$, which has a better data reduction ratio than $q = 5$, is satisfactory to achieve acceptable false positive and false negative rates.

Given $W = 8$, $L = 1024$, and $q = 4$, we need to determine proper correlation threshold coefficients ($K_l$) for

Table I: Practical evaluation of *q* (quantization bits)

|  | Data Reduction Ratio | SNR | Noise variance |
|---|---|---|---|
| $q = 3$ | 68 : 1 | 1 | $6.2e38$ |
| $q = 4$ | 40 : 1 | 1.4 | $1.2e38$ |
| $q = 5$ | 27 : 1 | 1.9 | $3.5e37$ |

SNR is determined by calculating the ratio of signal power to the noise power.

Table II: Theoretically Proper values for $K_l$

| $K_{300}$ | $K_{400}$ | $K_{500}$ | $K_{600}$ |
|---|---|---|---|
| 4.2 | 4.6 | 5 | 5.2 |

various excerpt sizes (*l*). Fig. 5 represents the theoretical false positive and false negative probabilities determined by Relations (6) and (17) for different $K_l$ values. As the figure shows, for an excerpt of size *l*, increasing $K_l$ results in lower false positive and higher false negative probabilities. Therefore, a middle value of $K_l$ must be selected that achieves both low false positive and low false negative probabilities. As the figure shows, larger excerpts, which have lower false negative probabilities, give the opportunity to select a larger $K_l$ and achieve a lower false positive probability. Considering the results of Fig. 5, we selected the threshold coefficients shown in Table II and used them in the following experiments.

To evaluate empirical false positive and false negative rates, we digested the traffic trace by DSPAS using $W = 8$, $L = 1024$, and $q = 4$. As mentioned earlier, the digest achieved a data reduction ratio of about 40:1. Then, we digested the traffic trace using the previous methods with the same data reduction ratio. To evaluate the false positive rate, we extracted 10000 excerpts of different sizes, while all of them had appeared only once in the traffic. We queried each PAS for determining the carrier flow of each excerpt.

Regarding the false negative issue, none of the methods resulted in such an outcome. The previous methods are naturally not expected to produce a false negative since they are based on hash functions. DSPAS, by contrast, can make a false negative response in theory. However, as the theoretical analysis showed, the proper setting of the system parameters made it practically very improbable.

Fig. 6 represents the empirical false positive rate of previous works alongside the empirical and theoretical false positive rates of DSPAS. The empirical false positive rate is the ratio of falsely detected flows to the total number of traffic flows. While CBID does not result in any false positives at this data reduction ratio, other previous methods have a small false positive rate. The false positive rate of DSPAS is higher than the previous methods, especially for small excerpts. However, it quickly gets close to the other methods as the excerpt size increases, and as can be seen, the false positive rate of DSPAS is near to zero from 500-byte excerpts upward.

As Fig. 6 shows, the false positive rate of DSPAS that we observed in practice is higher than what is expected from the theoretical analysis. There are two reasons for it. First, the theoretical analysis is based on perfect Gaussian distributions while it may not be entirely correct in practice. The second reason, as discussed in the last part of the previous section, is that an excerpt may not entirely fall into a single transform block and consequently, the effective excerpt length may be smaller than the excerpt's actual length. As stated before, we use a transform size of $L = 1024$ words (8196 bytes). Increasing the transform size can decrease the probability that a transform block does not entirely include an excerpt. Furthermore, DCT coefficients of larger transform sizes tend more to a Gaussian distribution. On the other hand, a large transform size increases the false positive rate according to relation (6). Therefore, a large transform size is not proper for small traffic flows. We practically examined different transform sizes. Fig. 7 illustrates the impact of the transform size on the false positive rate for different excerpt sizes. As the figure shows, neither very small nor very large transform sizes are efficient choices. The transform sizes 512 and 1024 words result in efficient false positive rates. We selected the transform size of 1024 words which is the optimum choice for excerpts of sizes 400 and 500 bytes.

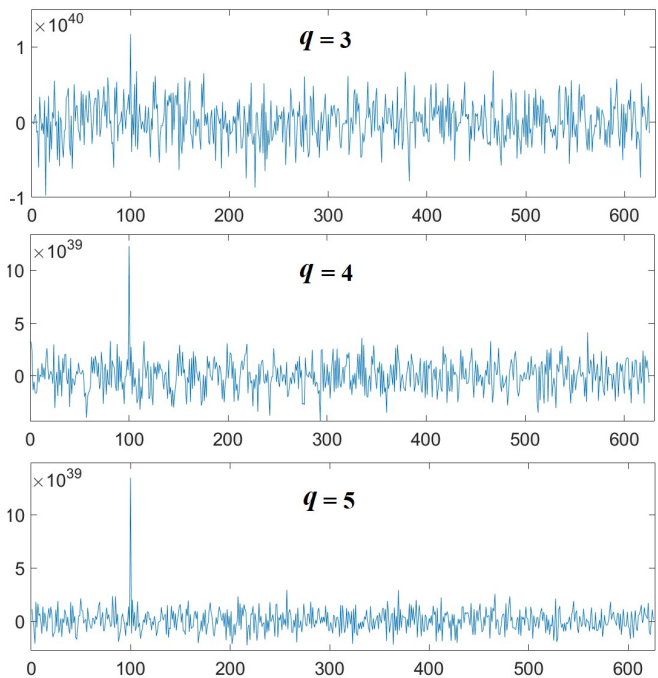

Fig. 4. Samples of correlation signal for detecting a 400-byte excerpt in digests obtained using different *q* values. The excerpt starts at point 100 of the recovered payloads.

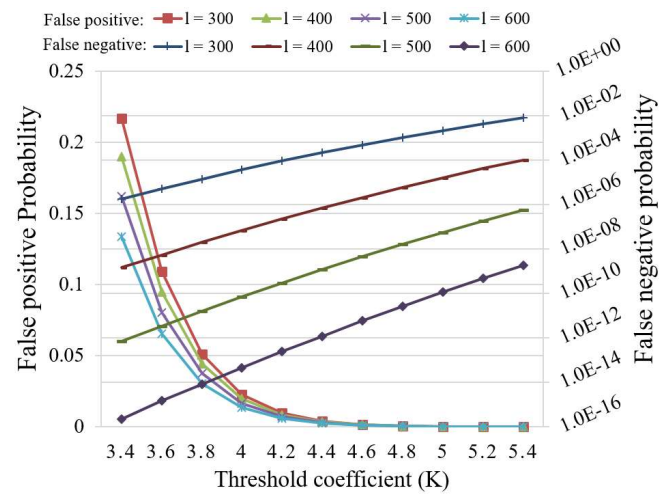

Fig. 5. Theoretical false positive and false negative probabilities calculated using relations 6 and 17 for different $K_l$ values. Increasing $K_l$ results in lower false positive and higher false negative probabilities

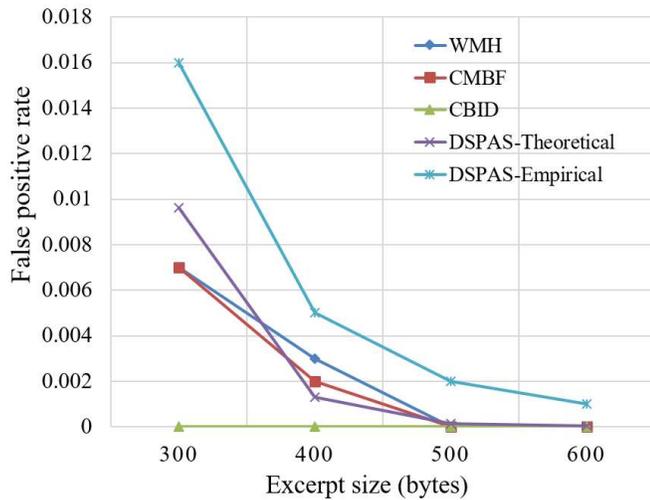

Fig. 6. Comparison of the false positive rate of payload attribution systems for simple (non-wildcard) queries. The false positive rate of each PAS is the number of falsely determined flows divided by the total number of traffic flows.

To evaluate wildcard queries and detection of similar strings, we used the 300-byte excerpts and marked a different number of bytes of them as wildcard bytes. The position of the wildcard bytes was randomly selected. At the same time, for each wildcard excerpt, we searched the traffic to ensure no other match exists for it except the original excerpt. Then, we queried CMBF and DSPAS for the wildcard excerpts. CMBF uses its character classification technique and comparing hash values to find the excerpts in the digest. On the other hand, DSPAS easily finds the strings similar to the queried excerpts using the correlation technique. The false positive rates are shown in Fig. 8. As can be seen, our method exhibits no increase in the false positive rate while the false positive rate of CMBF significantly increases with the number of wildcard bytes. As discussed before, the large number of simple queries which CMBF performs for wildcard queries is the cause of its high false positive rate. As the number of wildcard bytes increases linearly, the number of simple queries increases exponentially. Since each simple query has a chance of a false positive result (according to Fig. 6), a large number of them imposes a high false positive rate. On the contrary, the wildcard querying procedure of our method is the same as simple queries and is done in just a single query. It is also noteworthy that the false positive rate of CMBF **for wildcard queries** has not been evaluated by paper [12].

However, our method can result in a false negative response if the number of wildcard bytes increases. Fig. 9 represents the empirical false negative rate of DSPAS as a function of the excerpt size and the number of wildcard bytes. We did not observe any false negatives for excerpts of size 300 bytes composed of lower than 25 wildcard bytes. However, as expected, the false negative rate for wildcard queries is also related to the number of non-wildcard bytes, i.e. larger excerpts can tolerate more wildcard bytes.

Another important issue is the response time of wildcard queries. CMBF encounters extremely large response time when the number of wildcard bytes increases. The response time was so long that we could not get the result for wildcard queries comprising more than six wildcard bytes. It should be noted that the reported response time in CMBF's paper is only the response time of the appearance check step, and it does not reveal the total response time for answering a wildcard query. Indeed, the procedure for which they have reported a response time must be repeated for each traffic flow in the flow determination step. Paper [13] has

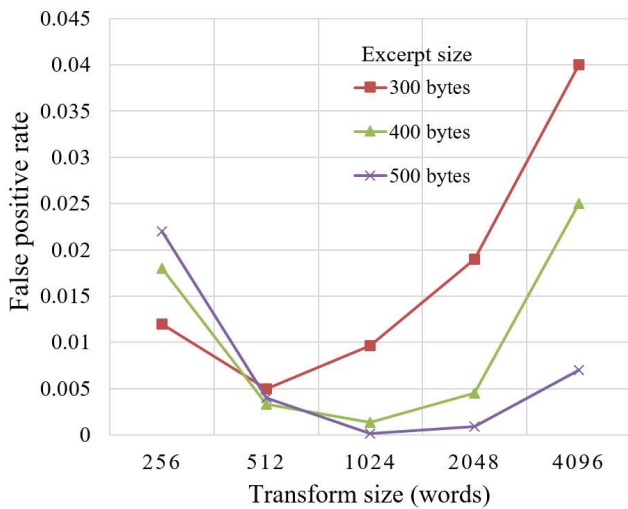

Fig. 7. The impact of the DCT transform size ($L$) on the false positive rate of DSPAS for various excerpt sizes.

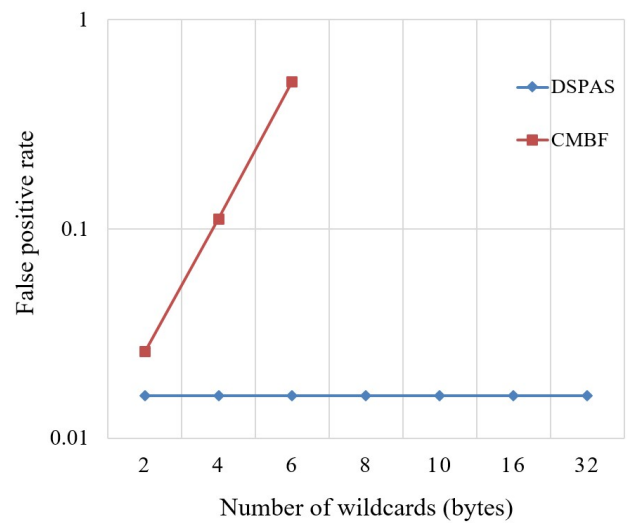

Fig. 8. Comparison of the false positive rate of DSPAS and CMBF in response to wildcard queries. The response time of CMBF for excerpts including more than 6 wildcard bytes was so long that we could not evaluate it. However, it seems that CMBF tends to a false positive rate equal to 1.

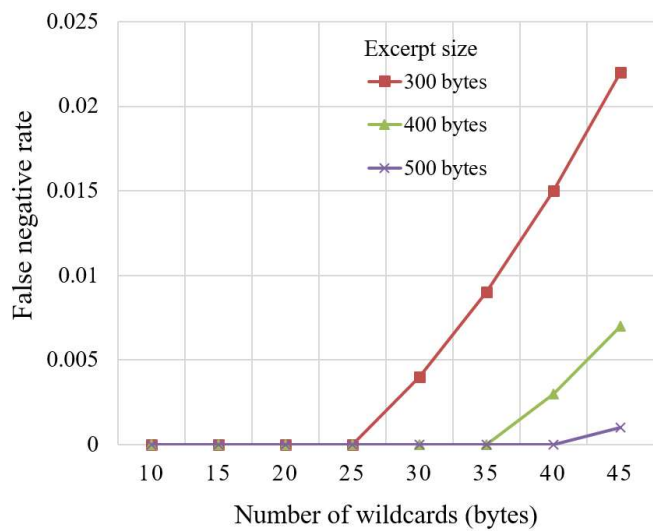

Fig. 9. Empirical false negative rate of DSPAS in response to wildcard queries.

TABLE III: Comparison of computation times

| Process | | Computation time (Minutes) | | | |
|---|---|---|---|---|---|
| | | WMH | CMBF | CBID | DSPAS |
| Digesting | | 6.0 | 5.1 | 5.9 | 27 |
| Simple query | | 0.12 | 0.10 | 0.20 | 22 |
| Wildcard query | WB = 2 | - | 0.9 | - | 22 |
| | WB = 4 | - | 18 | - | 22 |
| | WB = 6 | - | 311 | - | 22 |
| | WB = 8 | - | - | - | 22 |

WB: Number of wildcard bytes. The querying time is the computation time for querying one excerpt. Trace size is 4 GB.

comprehensively discussed this important issue.

Table III shows the processing time of digesting and querying for both types of excerpts using all the methods. We used an Intel Core i7-3770K CPU with 16 GB of RAM for this experiment. As can be seen in the table, since the computation complexity of DSPAS is more than the previous methods, its processing time for digesting and querying for simple excerpts is larger than the processing time of the other methods. However, when it comes to wildcard queries, DSPAS exhibits a more reasonable response time. While the querying time of CMBF increases exponentially with the number of wildcard bytes, the querying time of DSPAS is constant and suitable for any number of wildcard bytes.

## VI. Conclusion

In this paper, we presented a DSP-based approach to one of the important problems of network incident investigations, the payload attribution. For the first time in the literature, we used simple and widespread DSP techniques and showed that the approach has significant potential as an investigative method. We used the transform coding method for digesting network traffic. Thanks to the small size of the digest, the storage cost of network traffic archiving decreases considerably. Moreover, the digest preserves the privacy of users. We showed that the proposed method answers to query for excerpts of traffic with a bounded false positive rate. Although our method does not improve the false positive rate for simple queries in comparison to the previous works, it results in significantly better performance for wildcard queries in terms of the false positive rate and response time. While the previous methods are useless for wildcard queries because of their high false positive rate and excessively long response time, our approach can process the queries with a low false positive rate and a reasonable response time. Moreover, our method can easily detect strings similar to a queried excerpt. As a result, a malicious insider cannot easily evade the system. As future works, we will investigate how to improve the data reduction ratio of DSPAS. A possible approach to achieve a perfect PAS may be to combine the method of previous works, i.e. Bloom filters, with the method of this paper. We are also going to work on an efficient hardware implementation for real-time traffic digesting. Alleviation of the processing cost and trying other transforms such as the DFT and wavelets are also other interesting topics for future works.

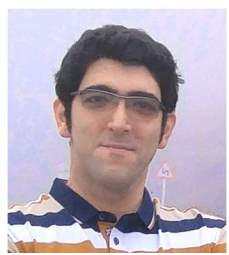

**S. Mohammad Hosseini** obtained his M.Sc. degree in Computer engineering from Sharif University of Technology, Tehran, Iran, in 2014. During his M.Sc. thesis, he has designed and implemented various high-performance network appliances based on FPGAs, such as a 40 Gb/s network equipment tester and high-bandwidth network traffic recorder and player. He is currently a Ph.D. student of Computer engineering at Sharif University of Technology, and his dissertation is on digesting network traffic for forensic applications. In addition to network forensics, his research interest areas include computer networks, computer architecture, and high-performance computing.

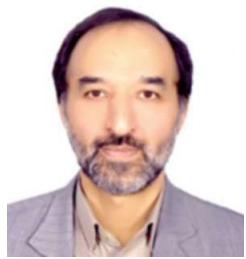

**Amir Hossein Jahangir** obtained his PhD. in Industrial informatics from Institut National des Sciences Appliquées, Toulouse, France in 1989. Since then, he has been with the Department of Computer Engineering, Sharif University of Technology, Tehran, Iran, and has served during his career as the Head of the Department, Head of Computing Center, and is now an associate professor and the director of Network Evaluation and Test Laboratory, an accredited internationally recognized laboratory in the field of network equipment test. His fields of interest comprise network equipment test and evaluation methodology, network security, and high-performance computer architecture.

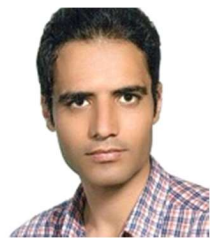

**Mehdi Kazemi** obtained his B.Sc. in Biomedical engineering from Isfahan University, Isfahan, Iran. He obtained his Master's degree in Computer engineering from Sharif University of Technology, Tehran, Iran, in 2014. His research interests include digital signal processing and hardware implementation of signal processing systems on FPGAs. He has participated in the design and implementation of various medical equipment.